\documentclass[fleqn,usenatbib]{mnras}
\usepackage[english]{babel}
\usepackage{amsmath, verbatim}
\usepackage[multidot]{grffile}
\usepackage{lineno}
\usepackage{natbib}
\usepackage{multirow}
\usepackage{array}
\usepackage{pifont}

\usepackage{xcolor}
\usepackage{booktabs}
\usepackage{cuted}
\usepackage{upgreek}
\usepackage{csquotes}
\usepackage{graphbox,graphicx}
\usepackage{subfig}
\usepackage{graphicx}
\usepackage{ulem}

\newcommand\numberthis{\addtocounter{equation}{1}\tag{\theequation}}

\begin{document}

\title{Fast emulation of two-point angular statistics for photometric galaxy surveys}
\author[M. Bonici et al.]{
Marco Bonici$^{1}$,
Luca Biggio$^{2}$,
Carmelita Carbone$^{1}$,
Luigi Guzzo$^{3,4}$
\\
$^{1}$INAF-IASF Milano, Via Alfonso Corti 12, I-20133 Milano, Italy\\
$^{2}$Data Analytics Lab, Institute of Machine Learning, Department of Computer Science, ETH Z\"urich, Switzerland\\
$^{3}$Dipartimento di Fisica ``Aldo Pontremoli", Universit\`a degli Studi di Milano, \& INFN, Sez. di Milano, Via Celoria 16, I-20133 Milano, Italy\\
$^{4}$INAF, Osservatorio Astronomico di Brera, 
Milano, Italy\\
}

\maketitle
\begin{abstract}
We develop a set of machine-learning based cosmological emulators, to obtain fast model predictions for the $C(\ell)$ angular power spectrum coefficients characterising  tomographic observations of galaxy clustering and weak gravitational lensing from multi-band photometric surveys (and their cross-correlation).
A set of neural networks are trained to map cosmological parameters into the coefficients, achieving a speed-up $\mathcal{O}(10^3)$ in computing the required statistics for a given set of cosmological parameters, with respect to standard Boltzmann solvers, with an accuracy better than 0.175\% ($<$0.1\% for the weak lensing case).  
This corresponds to $\sim$ 2\% or less of the statistical error bars expected from a typical Stage IV photometric surveys. Such overall improvement in speed and accuracy is obtained through (\textit{i}) a specific pre-processing optimisation,  ahead of the  training phase,  and (\textit{ii}) a more effective neural network architecture, compared to previous implementations. 
\vspace{1.2cm}
\end{abstract}

\section{Introduction}
The golden age of cosmology is continuing. The discovery of the accelerated expansion of the universe~\citep{perlmutter_measurements_1999, riess_observational_1998} led to the consolidation of the $\Lambda$CDM model, which can now be convincingly defined as the standard model of cosmology. 

We are currently  witnessing a period of fast developments, from the theoretical and observational points of view, aimed at understanding its details and ingredients.
In fact, the standard model is capable to successfully predict virtually all early and late universe observations to high accuracy using a restricted number of parameters, although some residual, perhaps telltale, tensions should not be overlooked \citep[\textit{e.g.}][]{bernal_trouble_2016}. However, the very nature of this model opens a number of fundamental questions which await a satisfactory answer, as in particular the nature of its dark sector components, Dark Matter and Dark Energy.

Together with CMB anisotropy observations \citep[\textit{e.g.}[]{aghanim_planck_2020}, one of the key pillars supporting the standard model is represented by measurements derived from galaxy surveys. On the one hand, statistical measurements of {\it galaxy clustering} (GC hereafter) in 3D (when spectroscopic redshifts are available) or 2D (\textit{i.e.} in projection on the sky) provide us with an \textit{indirect} probe (through luminous tracers) of the distribution of matter density fluctuations on different scales and at different epochs~\citep{feldman_power_1994, yamamoto_measurement_2006}. On the other hand, photometric imaging surveys allow us, under certain conditions, to measure galaxy shapes and use their shear, due to {\it weak gravitational lensing} (WL hereafter) by foreground galaxies, to assess \textit{directly} the matter density field, providing us with the ability to perform a {\it tomography} of the large-scale matter distribution in the Universe~\citep{hu_power_1999}. Combinations of photometric 2D-GC angular correlation function, WL correlation function, and GC-WL cross-correlation between galaxy positions and shapes (3x2pt  statistic hereafter), together with the spectroscopic 3D-GC, will enable unprecedented precision and accuracy to be achieved in the estimate of cosmological parameters.

We are on the verge of a new era for galaxy surveys, with the next generation of experiments that aim at solving the mysteries of the standard cosmological model. This is the case, in particular, of the DESI ground-based spectroscopic survey \citep{abbott_dark_2018-1}, recently started at KPNO, and the ESA Euclid mission \citep{laureijs_euclid_2011}, a combined imaging and spectroscopic experiment, currently scheduled for launch in 2023.  The multi-band, multi-epoch photometric LSST survey by the Vera C. Rubin telescope \citep{weltman_fundamental_2020}, and the future Nancy Roman space telescope \citep{dore_wfirst_2019}, will provide further unprecedented photometric and spectroscopic galaxy data, contributing to the accumulation of an enormous amount of information. The Euclid mission, in particular, will play a special role in this respect, with its ability to simultaneously deliver measurements of both spectroscopic and photometric galaxy clustering together with weak gravitational lensing (and further additional probes, like, \textit{e.g.} galaxy clusters -- see \citet{blanchard_euclid_2020}).

Such an avalanche of new experimental data, naturally calls for a comparable quantum leap in the efficiency of our analysis tools. 
Each of these new data sets will be massive on its own. Additionally, the combination and cross-correlation of different probes (GC and WL, but also CMB anisotropies, void statistics, clusters, etc.) will leverage their information content, while providing firmer control over systematic errors.
Besides the theoretical challenge of accurately modelling the data down to small, nonlinear scales~\citep{carrasco_effective_2012, cataneo_road_2019, euclidcollaboration_euclid_2021, angulo_bacco_2021}, we will be facing a serious \textit{computational} challenge: cosmological analyses are based on Markov Chain Monte Carlo (MCMC) techniques, which require theoretical predictions to be computed as many as $10^4$ to even $10^6$ times, with the execution of so-called Boltzmann solvers at each step of the chains.  
The most widely used of such codes are \texttt{CAMB}~\citep{lewis_efficient_2000} and \texttt{CLASS}~\citep{lesgourgues_2011}, which compute the evolution of linear perturbations in about $\approx$ 0.5 seconds (7 seconds when massive neutrinos are included) on a standard laptop\footnote{Test performed using a laptop with a i7-10510U CPU.}. Such basic figures clearly show that performing a full analysis of cosmological datasets requires either a very long time or the access to high-performance computing facilities.

To overcome this issue, codes called {\it emulators}, capable of reproducing the results of Boltzmann solvers, but interpolating from a coarser grid of ``pivot" models, have been recently developed, achieving significant speed-up. The first of such codes, based on polynomial regression, were used in the analysis of CMB data \citep{jimenez_fast_2004, fendt_pico_2007}. A different approach was taken in~\cite{albers_cosmicnet_2019}, where a neural network was trained to replace the most expensive parts of the \texttt{CLASS} Boltzmann solver.

More recently, emulators were developed to produce model predictions for large-scale structure two-point statistics
\citep{manrique-yus_euclid-era_2019, mootoovaloo_parameter_2020, mootoovaloo_kernel-based_2021, arico_accelerating_2021, mancini_itcosmopower_2021}. These codes allow us to emulate either the spatial 3D matter power spectrum, $P_\mathrm{mm}(k)$, in real space, or the angular power spectrum, $C(\ell)$, \textit{i.e.} the coefficients of the spherical harmonic decomposition of the 2D correlation function, which quantifies density fluctuations as a function of the angular scale alone. The latter is customary when describing CMB anisotropies, weak lensing measurements or any statistics measured in projection on the sky, as also 2D galaxy clustering within ``tomographic slices".

In this paper, we describe the development of a set of three new emulators to derive fast model predictions for the $C(\ell)$ associated to three different observables. In particular, we focus on the 3x2pt statistics associated to two of the main cosmological probes of next generation surveys, GC and WL, together with their cross-correlation. Specifically, the cross-correlation of the tiny and coherent image distortions of background galaxies with the angular 2D galaxy distribution is of special interest and will be among the key observables of, for instance, the Euclid and LSST galaxy surveys.

We adopt a machine learning approach, building and training a set of neural networks to map cosmological parameters into the 3x2pt angular correlation coefficients $C(\ell)$. Neural networks are becoming more and more popular in many areas of science and engineering. Much of their success stems from their universal approximation properties, \textit{i.e.} their ability to approximate any continuous function, provided enough data and an appropriate number of {\it hidden layers} \citep{cybenko_approximation_1989, hornik_multilayer_1989} are combined.  Recent years have witnessed the emergence of powerful neural networks in the form of different architectures tailored to the specific task being performed. Such models are typically built by stacking multiple layers of neurons, the so-called {\it deep neural networks}.

The main advantage of a machine learning approach to cosmological inference is the significant speed-up of the full analysis. In fact, concerning the computation of $P_\mathrm{mm}(k)$, in the standard approach the hierarchy of coupled Einstein-Boltzmann equations needs to be solved, and the solution is numerically integrated to obtain the desired quantities. Emulators skip these steps, with a gain of orders of magnitude in speed. The gain is even more pronounced when emulators are applied directly to cosmological observables and tomographic analyses, which are our main interest here~\citep{hu_power_1999}: for example, in an analysis involving both GC, WL, and their cross-correlation, with 10 tomographic redshift bins the value of 210 different $C(\ell)$ needs to be computed, for each multipole $\ell$. This is computationally very expensive if performed via standard approaches, especially within MCMC techniques for cosmological parameter inference. Thus, a direct emulation of the angular power spectra would provide a huge speed-up of the full analysis.

As discussed in the following, our implementation allows us to emulate the 3x2pt angular power spectrum coefficients $C(\ell)$ for a typical Stage-IV galaxy survey, with a speed-up $\mathcal{O}(10^3)$ with respect to the standard approach, while preserving an accuracy of about $0.175\%$ over the whole emulated range of angular scales.  Besides their speed and accuracy, our emulators can be easily extended to  larger parameter spaces. These figures compare very favourably to similar works in the literature.

Of course, the gain in terms of time needed for cosmological parameter inference comes at the price of data generation and training processes, which can be also computationally expensive. 
The overall balance, however, remains positive,
given the massive computational costs needed to generate standard MC chains and the property of deep neural networks to be easily generalized to high-dimensional parameter spaces. In the present work, emulators are treated as black-box functions and do not incorporate any \textit{explicit} prior knowledge of the underlying physics. They learn the mapping between input parameters and output correlation coefficients in an entirely data-driven fashion. 

This paper is structured as follows.
In Section~\ref{sec:setup} we explain how we produced and preprocessed the training dataset, the neural network architecture and the training procedure. In Section~\ref{sec:performance} we test the precision and speed of our emulators. In Section~\ref{sec:comparison} we make a comparison with the emulators present in literature. We then summarise our results and conclude in Section~\ref{sec:conclusion}.

\section{Methodology}
\label{sec:setup}

\subsection{Theoretical Background}
\label{sec:theory}
We give here a brief overview of the theory underlying two-point statistics used in the analyses of LSS surveys, using a notation similar to~\cite{blanchard_euclid_2020}. Throughout the paper we consider as reference background model the simplest generalization of the standard $\Lambda$CDM model, in which dark energy is assumed to be a dynamical quantity with Equation of State (EoS) described by the Chevalier~-Polarski-~Linder parameterisation: $w(a) = w_0-w_a(1-a)$~\citep{chevallier_accelerating_2001, linder_exploring_2003}. We refer to this background model as a flat ``$w_0w_a$CDM" model.

As mentioned above, our aim is to emulate three specific sets of two-point angular statistics involving galaxy clustering and weak lensing, accounting also for the correlation between different tomographic redshift bins: (\textit{i}) the auto-correlation of angular galaxy positions in a {\it photometric} sample; (\textit{ii}) the cross-correlation of the lens galaxy positions with the estimated shear, and (\textit{iii}) the auto-correlation of the shear field itself. We describe these quantities in terms of $C(\ell)$ angular power spectra, \textit{i.e.} the coefficients of the spherical harmonic decomposition of the above correlation functions, with the multipole order, $\ell$, related to the inverse of the angular scale in configuration space.

In the standard approach, each $C(\ell)$ corresponds to a projection of the nonlinear matter power spectrum, $P_\mathrm{mm}\left(k, z\right)$, weighted with the window functions $W^A(z)$ and $W^B(z)$ associated to the probes $A$ and $B$ (here $\gamma\equiv{\rm WL}$ and $g\equiv{\rm GC}$, respectively) where for $A=B$ we recover the auto-power spectrum of a single probe. Under the Limber approximation~\citep{limber_analysis_1953, loverde_extended_2008}, these coefficients can be expressed as
\begin{equation}
	\label{eq:cl_limber}
	C_{ij}^{AB}(\ell) = \frac{c}{H_0} \int_{z_\text{min}}^{z_\text{max}} \mathrm{d} z \frac{W_{i}^{A}(z) W_{j}^{B}(z)}{E(z) r^{2}(z)} P_\mathrm{mm}\left(k_\ell(z), z\right) ,
\end{equation}
where $E(z)=H(z)/H_0$ is the dimensionless Hubble parameter, ${z_\text{min}}$ and ${z_\text{max}}$ are the minimum and maximum redshifts of the survey, $r(z)$ is the comoving distance, $i$ and $j$ label the tomographic redshift bins, and $k_\ell$ is the wavenumber,  which is connected to the multipole order, $\ell$, as this
\begin{equation}
    k_\ell(z)=\frac{\ell+1/2}{r(z)}.
\end{equation}
In the adopted flat $w_0w_a$CDM cosmology, $E(z)$ reads
\begin{equation}
E(z)=\sqrt{\Omega_{\mathrm{m}, 0}(1+z)^{3}+\Omega_{\mathrm{DE}, 0}(1+z)^{3\left(1+w_{0}+w_{a}\right)} \mathrm{e}^{-3 w_{a} \frac{z}{1+z}}},
\end{equation}
and the comoving distance is
\begin{equation}
r(z)=\frac{c}{H_{0}} \int_{0}^{z} \frac{\mathrm{d} z}{E(z)}.
\label{eq:comoving_distance}
\end{equation}
Still following \citet{blanchard_euclid_2020}, the WL window function is given by
\begin{equation}
W_{i}^{\gamma}(z) = \frac{3}{2} \left( \frac{H_0}{c}\right) ^2 \Omega_M (1+z) r(z)\tilde{W}_i^\gamma(z)\,,
\label{eq:w_lensing}
\end{equation}
where $\tilde{W}_i^\gamma(z)$ is the so-called \textit{lensing efficiency}
\begin{equation}
\tilde{W}_i^\gamma(z)=\int_{z}^{z_{\max }} \mathrm{d} z^{\prime} n_{i}^g\left(z^{\prime}\right)\left(1-\frac{r(z)}{r\left(z^{\prime}\right)}\right),
\label{W^gamma}
\end{equation}
with $n_i^g(z)$ being the normalized surface galaxy number density in the considered $i$-th tomographic bin (see below for details).

Unlike WL, the GC selection function needs to account also for the galaxy bias $b_i^g(z)$ (see \cite{desjacques_large-scale_2018} for a review),
\begin{equation}
    \label{eq:w_galaxyclustering}
	W_{i}^{g}(z) = \frac{H(z)}{c} n^g_{i}(z) b^g_i(z).
\end{equation}
We use the phenomenological galaxy bias model introduced in~\cite{tutusaus_euclid_2020},
\begin{equation}
    b(z)=A+\frac{B}{1+\exp [-(z-D) C]},
    \label{eq:bias_model}
\end{equation}
with $A = 1.0$, $B = 2.5$, $C = 2.8$, and $D = 1.6$, as obtained through a fit to the Euclid flagship simulation\footnote{Euclid Collaboration, in preparation.}.

Here we model the true galaxy distribution as
\begin{equation}
	n^g(z) \propto \left(\frac{z}{z_{0}}\right)^{2} \exp \left[-\left(\frac{z}{z_{0}}\right)^{3 / 2}\right]\,,
	\label{eq:galaxy_density}
\end{equation}
where $z_0=z_m/\sqrt{2}$ is the median redshift of the distribution 
normalized as to obtain an integrated galaxy surface density $\bar{n}^g$.
Here we assume the same specifications as for the Euclid photometric survey~\citep{laureijs_euclid_2011}, \textit{i.e.} $z_m=0.9$ and $\bar{n}^g=30 \operatorname{arcmin}^{-2}$.

To account for photometric redshift errors, we convolve the true galaxy surface density, $n^g(z)$, with the following double Gaussian kernel describing the probability
that a galaxy with redshift $z$ has a measured redshift $z_p$~\citep{kitching_cosmological_2009}
\begin{align*}
p(z|z_p) \, &= \, \frac{1-f_{\mathrm{out}}}{\sqrt{2 \pi} \sigma_{\mathrm{b}}(1+z)} \exp \left\{-\frac{1}{2}\left[\frac{z-c_{\mathrm{b}} z_{\mathrm{p}}-z_{\mathrm{b}}}{\sigma_{\mathrm{b}}(1+z)}\right]^{2}\right\}\\
&+\frac{f_{\mathrm{out}}}{\sqrt{2 \pi} \sigma_{\mathrm{o}}(1+z)} \exp \left\{-\frac{1}{2}\left[\frac{z-c_{\mathrm{o}} z_{\mathrm{p}}-z_{\mathrm{o}}}{\sigma_{\mathrm{o}}(1+z)}\right]^{2}\right\} \numberthis
\label{eq:pzpz}\,.
\end{align*}
This is quite a flexible parameterization, which includes both multiplicative and additive bias in the redshift determination, accounting for a fraction $f_{\text {out }}$ of galaxies with incorrect estimate of redshift; the values of the coefficients are given in Tab.~\ref{tab:pzpz}.
\begin{table}
	\centering
	\begin{tabular}{  c | c| c |c |c |c |c  }
	\toprule
		{$c_{b}$} & {$z_{b}$} & {$\sigma_{b}$} & {$c_{o}$} & {$z_{0}$} & {$\sigma_{o}$} & {$f_{\text {out }}$} \\
		\midrule
		1.0 & {0.0} & {0.05} & {1.0} & {0.1} & {0.05} & {0.1}  \\  
		\bottomrule	
	\end{tabular}
	\caption{Parameters used in the photometric redshift distribution $p(z|z_p)$ of Eq.~\eqref{eq:pzpz} and Fig.~\ref{fig:ngz}.}
	\label{tab:pzpz}
\end{table}
Finally, the number density in the $i$-th bin, entering Eqs. ~\ref{eq:w_galaxyclustering}-\ref{W^gamma}, is \begin{equation}
n_{i}^g(z)=\frac{\int_{z_{i}^{-}}^{z_{i}^{+}} \mathrm{d} z_{\mathrm{p}} n^g(z) p_{\mathrm{ph}}\left(z_{\mathrm{p}} \mid z\right)}{\int_{z_{\min }}^{z_{\max }} \mathrm{d} z \int_{z_{i}^{-}}^{z_{i}^{+}} \mathrm{d} z_{\mathrm{p}} n^g(z) p_{\mathrm{ph}}\left(z_{\mathrm{p}} \mid z\right)},
\end{equation}
where $z_i^+$ and $z_i^-$ are the redshift values that define the $i^{th}$ tomographic bin. Here we use ten equi-populated redshift bins, so that the number density in each bin is simply $\bar{n}^g_i=\bar{n}^g/10$ and the boundaries of the bins are:
\begin{equation*}
    z_{i}=\{0.001,0.42,0.56,0.68,0.79,0.9,1.02,1.15,1.32,1.58,2.5\}.
\end{equation*}

\begin{figure}
\includegraphics[width=0.48\textwidth]{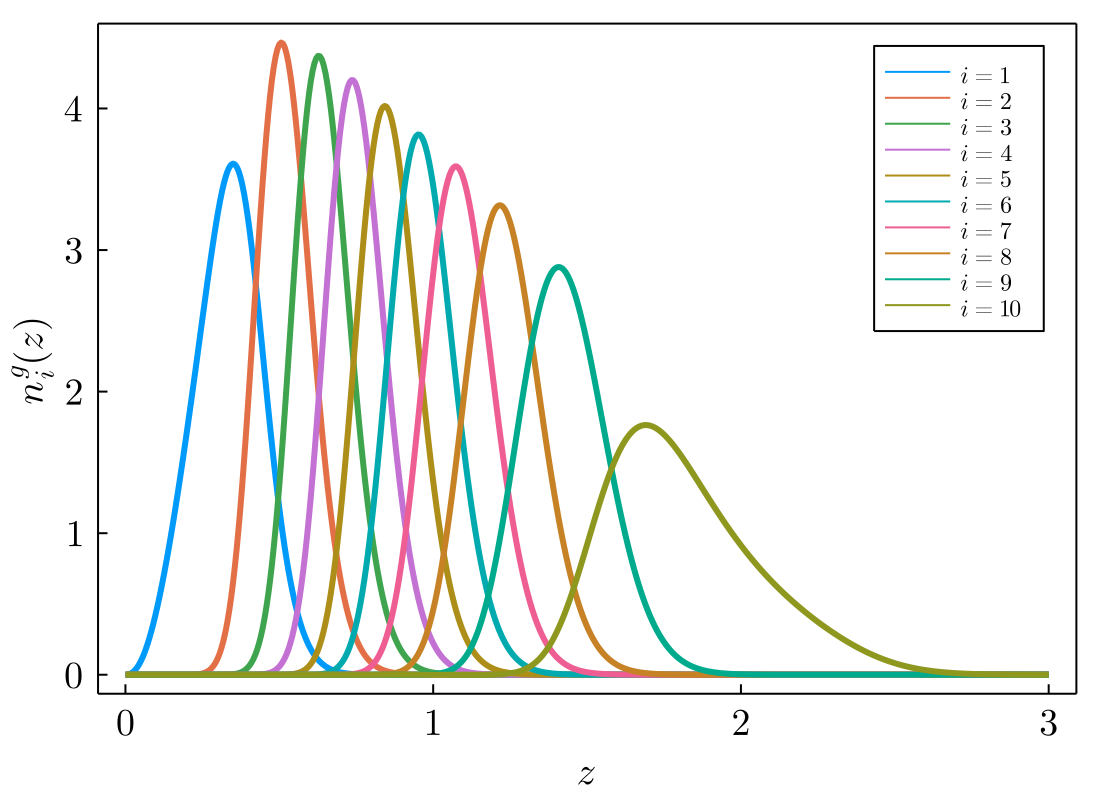}

\caption{Normalized galaxy density distributions in tomographic redshift bins, accounting for the photometric errors given in Tabl.\ref{tab:pzpz}, used in this work.}
\label{fig:ngz}
\end{figure}

\begin{figure*}
\includegraphics[width=1\textwidth]{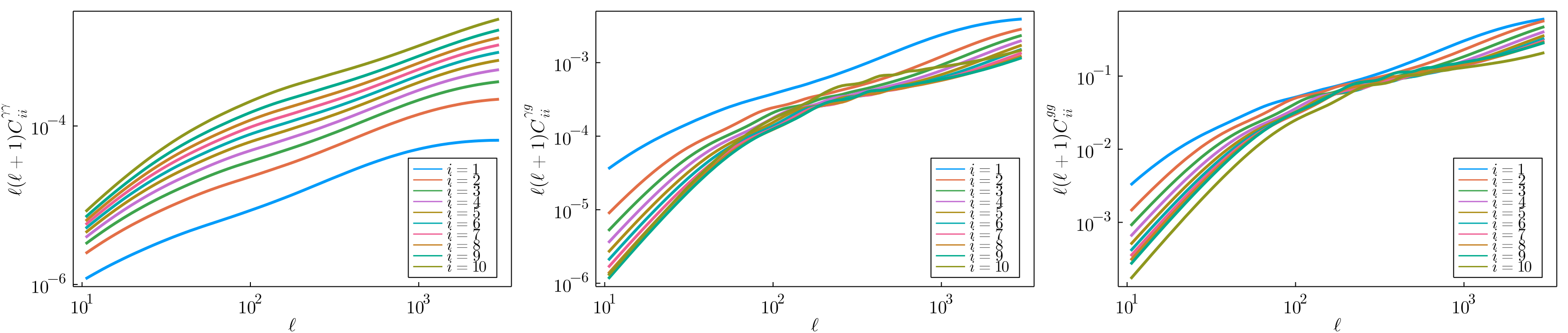}

\caption{Theoretical predictions for, respectively, lensing auto-correlation, galaxy-galaxy lensing cross-correlation, galaxy clustering auto-correlation. The power-spectrum was generated using \texttt{CLASS} in the central cosmology of Table~\ref{tab:parameters_range}, while the $C(\ell)$'s have been evaluated using \texttt{CosmoCentral.jl} .}
\label{fig:theory_predictions}
\end{figure*}

\subsection{Data set generation}
In this subsection, we describe both the dataset used in our analysis and the operations performed before the training phase.
The $C(\ell)$ were evaluated using our code named \texttt{CosmoCentral.jl}\footnote{The code is publicly available on \href{https://github.com/marcobonici/CosmoCentral.jl}{GitHub}~(Bonici \& Carbone, in preparation).}, a cosmological analysis package developed in \texttt{Julia}\footnote{\texttt{Julia} is a just-in-time compiled language, developed for high-performance numerical computations. Here, \texttt{Julia} version 1.6.3 was used.}.
The linear matter power-spectrum was obtained from \texttt{CLASS} and corrected for nonlinear evolution using the revised \texttt{halofit} recipe from~\cite{takahashi_revising_2012}.

The parameter hyperspace described in Table~\ref{tab:parameters_range} was then sampled using 200 logarithmically-spaced multipoles $\ell$, while  10,000 combinations of cosmological parameters were generated according to the Latin Hypercube Sampling (LHS) algorithm~\citep{latin_hypercube}. This prior was chosen to closely mimic that of \texttt{EuclidEmulator2}~\citep{euclidcollaboration_euclid_2021}.
\begin{table}
\centering
\begin{tabular}{|c |c |c |} 
 \hline
 Parameter & Range & Centre\\ [0.5ex] 
 \hline\hline
 $\Omega_M$ & [0.29, 0.35] & 0.32\\ 
 $\Omega_B$ & [0.04, 0.06] & 0.05\\
 $h$ & [0.61, 0.73] & 0.67\\
 $n_s$ & [0.92, 1.0] & 0.96\\
 $\sigma_8$ & [0.6, 1.0] & 0.8\\
 $w_0$ & [-1.3, -0.7] & -1.0\\
 $w_a$ & [-0.7, 0.7] & 0.0\\
 $\ell$ & [10, 3000] & --\\[1ex] 
 \hline
\end{tabular}
\caption{Ranges of cosmological parameters used for the training dataset generation, chosen as to be similar to those of EuclidEmulator2~\citep{euclidcollaboration_euclid_2021}.}
\label{tab:parameters_range}
\end{table}

\subsection{Neural network emulator}
Previous emulators~\citep{mootoovaloo_parameter_2020, mootoovaloo_kernel-based_2021} were based on Gaussian processes, which have a neat probabilistic interpretation and allow an easy estimate of the uncertainty associated with their predictions. However, they are not ideal for handling big data sets. The main disadvantage is the need of inverting an $N\times{N}$ covariance matrix to calculate the posterior distribution, with $N$ being  the length of the data vector in the training dataset.

Neural networks, instead, have proven to scale favourably with the size of the training data. This is a crucial point since, when feasible, by increasing the size of the training dataset we can improve the precision and accuracy of the emulator.

\begin{figure}
	\centering
	\includegraphics[width=0.9\linewidth]{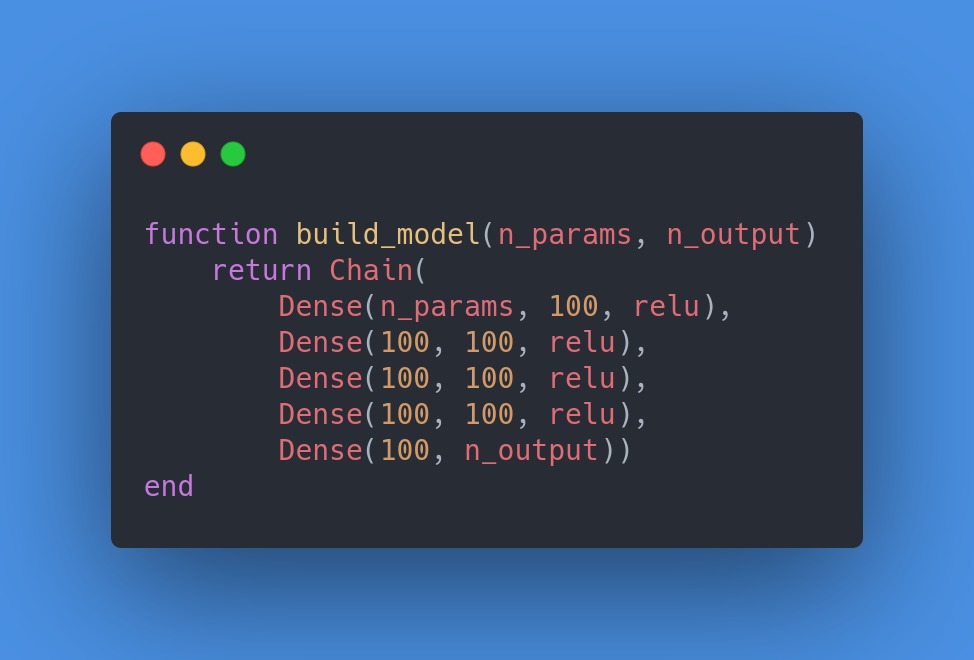}
    \caption{Architecture definition of our neural network emulators. \texttt{n\_output} changes accordingly to the emulator considered (55 and 100, respectively, for the auto-correlation and cross-correlation functions). }
    \label{fig:nn_arch}
\end{figure}

We address our problem as a classic \textit{regression task}: given the value of the input parameters (the cosmological parameters and the chosen multipole $\ell$), a neural network is trained to output the set of angular correlation coefficients, $C_{ij}^{AB}(\ell)$, for each of the three probes described in Sec.~\ref{sec:theory}.
The three neural networks are developed using \texttt{Flux.jl}, a \texttt{Julia} machine learning library~\citep{innes_fashionable_2018}.
The resulting \texttt{Julia} emulators can be easily wrapped and used in established \texttt{Python} data-analysis frameworks~\citep{torrado_cobaya_2021, brinckmann_montepython_2018}.
The chosen architecture consists of four hidden layers with 100 neurons each, and a Rectified Linear Unit (ReLu) activation function~\citep{nwankpa_relu}, as presented in Figure~\ref{fig:nn_arch}. The input layer takes 8 features (corresponding to the cosmological parameters and the multipole $\ell$). 
The output layer differs from that of similar emulators in the literature~\citep{mootoovaloo_parameter_2020, manrique-yus_euclid-era_2019}, which typically compute the $C(\ell)$ for a specific tomographic combination, $ij$, and in a given multipole range (or even at a specific multipole value). In our case, instead we output 55 coefficients for the auto-power spectra, $C_{ij}^{AA}(\ell)$, and 100 coefficients for the cross-power spectra, $C_{ij}^{AB}(\ell)$, \textit{i.e.}
we output at once \textit{all combinations}, $(i,j)$, of tomographic indices, with $i,j=1...10$, for the required multipoles.
This approach has several advantages:
\begin{itemize}
    \item Both the training and the predictions are \textit{fast}, since we use only three emulators, rather than the several tens required, \textit{e.g.}, in~\cite{manrique-yus_euclid-era_2019} to emulate the photometric 3x2pt.
    \item $C_{ij}^{AB}(\ell)$ are not \textit{un}correlated between different tomographic bins. We take advantage of this correlation by emulating all tomographic combinations at once.
    \item Our emulator suite is \textit{lightweight}, occupying less than 1 MB of disk storage.
\end{itemize}
The architecture of the emulators is presented in Figure~\ref{fig:nn_arch}. 

A step of the utmost importance before training is the \textit{preprocessing} of the training dataset, which helps the neural network to better capture the features it has to learn, both reducing the training time and increasing the neural network performance~\citep{ioffe_batch_2015}. Specifically, we first perform the following transformation
\begin{equation}
C_{ij}^{AB}(\ell) \longrightarrow \log( \ell(\ell+1))C_{ij}^{AB}(\ell),
\end{equation}
and then normalize the resulting values within the $[0,1]$ interval. We empirically verified the importance of this step: compared to a preliminary attempt, in which we tried to directly emulate the $C_{ij}^{AB}(\ell)$, this pre-processing yields a considerable improvement.
Regarding the training phase, we chose the ADAM optimizer~\citep{kingma_adam}, with a learning rate of $3\cdot10^{-4}$ and a batch-size of 256; the training lasted for 1,000 epochs. $80\%$ of the data set was used for training and $20\%$ for the validation. We used a mean-square-error loss function, which was evaluated after each epoch both for the training and the validation data sets; when the test loss function decreases, the stored network coefficients are updated.

\section{Emulator performance}
\label{sec:performance}
We assess the speed of our emulators using \texttt{BenchmarkTool.jl}, which runs the emulators several times and provides an accurate estimate of execution times. In this test, we emulated 30 multipoles, $\ell$, with the lensing auto-correlation emulator, obtaining a mean execution time of about 300 $\upmu$s. 

\begin{figure*}
\includegraphics[width=1\textwidth]{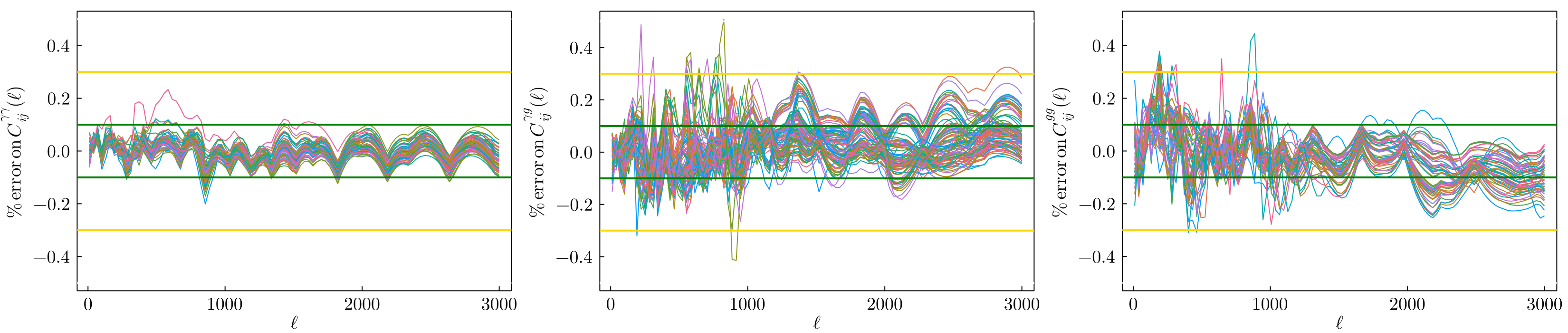}
\caption{The relative error in the $C(\ell)$ evaluation for the central cosmology defined in Table~\ref{tab:parameters_range}. The yellow and green lines represent, respectively, a relative difference of $0.3$ and $0.1\%$. The lensing emulator is the most precise, with an error smaller than $0.1\%$ for almost all evaluated coefficients. The superior performance of the WL emulator is due to the absence of the BAO wiggles, as can be seen in Figure~\ref{fig:theory_predictions}.}
\label{fig:residuals}
\end{figure*}

\begin{figure}
	\centering
	\includegraphics[width=0.9\linewidth]{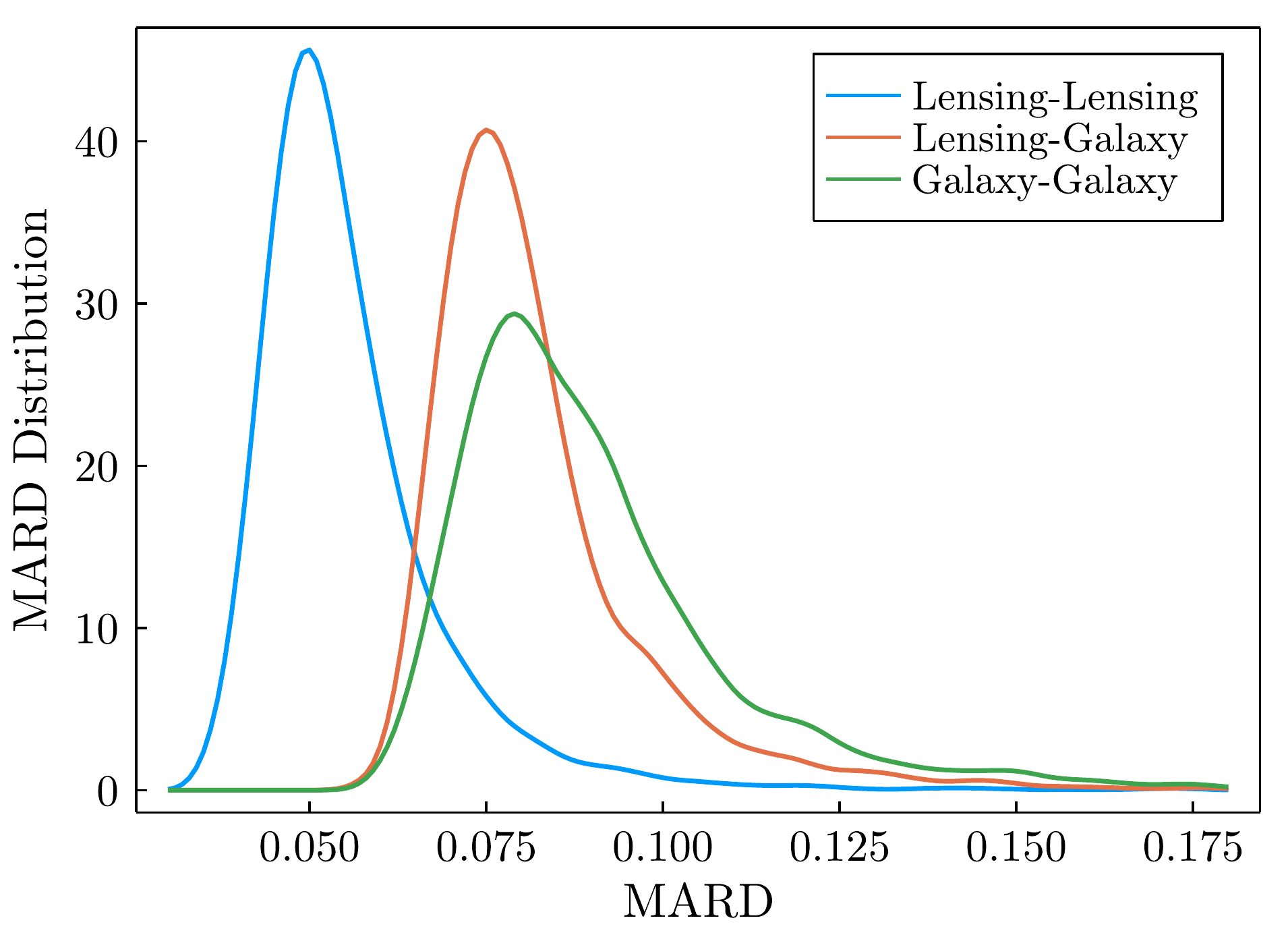}
    \caption{The distribution of the relative systematic errors of the three emulators, defined in Equation~\eqref{eq:MARD}, using a validation dataset of 4,000 cosmologies.}
    \label{fig:MARD}
\end{figure}

In Figure~\ref{fig:residuals}, we show the relative emulation error for the specific set of cosmological parameters corresponding to the centre of the emulated space, as reported in Table~\ref{tab:parameters_range}. Remarkably, the accuracy is very good over all multipoles considered: the residuals for the lensing auto-correlation are almost always below 0.1\%, while for the other two angular statistics they almost never exceed the 0.3\% threshold.
We also built a test dataset of 4,000 cosmologies distributed, according to the Latin Hypercube  Sampling, over the same range as the training dataset reported in Table~\ref{tab:parameters_range}.
To quantify the goodness of the emulators, we use the Mean Absolute Relative Difference (MARD) for each combination $\boldsymbol{\theta}$ of cosmological parameters present in the validation dataset, defined as
\begin{equation}
    \mathrm{MARD}^\mathrm{AB}(\boldsymbol{\theta})= \frac{100}{N_{C(\ell)}}\sum_{\ell=\ell_\text{min}}^{\ell_\text{max}}\sum_{i,j}\Bigg | 1-\frac{C_{ij,\text{NN}}^\mathrm{AB}(\boldsymbol{\theta})}{C_{ij,\text{GT}}^\mathrm{AB}(\boldsymbol{\theta})}\Bigg |,
    \label{eq:MARD}
\end{equation}
where $C_{ij,\text{NN}}^\mathrm{AB}$ is the neural network prediction,  $C_{ij,\text{GT}}^\mathrm{AB}$ is the ground truth, computed in the standard way (Eq.~\eqref{eq:cl_limber}), and $N_{C(\ell)}$ represents the number of $C_\ell$'s computed for each combination of cosmological parameters. This metric quantifies the \textit{relative} accuracy of the emulator. The resulting distribution is shown in Figure~\ref{fig:MARD}.
The important result is that the relative error of the predictions is well within $0.175\%$, with the lensing emulator, in particular, yielding an accuracy of $0.1\%$.
This higher accuracy, compared to the other two cases, is due to the negligible effect of Baryon Acoustic Oscillations (BAO) on $C_{ij}^\mathrm{\gamma \gamma}$, which make the weak lensing angular power spectra smoother and hence easier to emulate than $C_{ij}^\mathrm{gg}$ and $C_{ij}^\mathrm{\gamma g}$. In all cases, these errors are well below the typical uncertainties of nonlinear corrections to the matter power-spectrum: for example, \texttt{EuclidEmulator2}~\citep{euclidcollaboration_euclid_2021} and \texttt{BaccoEmulator}~\citep{angulo_bacco_2021}, have an accuracy of about $1-2\%$ up to $k=5\,h\text{Mpc}^{-1}$.
To estimate the significance of this level of systematic error with respect to the typical measurement uncertainty, we compare it to the expected cosmic variance $\sigma(\ell)$
\begin{equation}
    \sigma_{ij}^\mathrm{AB}({\ell})=\sqrt{\frac{C^\mathrm{AA}_{ii}({\ell}) C^\mathrm{BB}_{jj}({\ell})+\left(C^\mathrm{AB}_{ij}({\ell})\right)^{2}}{f_\mathrm{sky}(2 \ell+1)}}\, ,
\end{equation}
where $f_\mathrm{sky}$ is the fraction of surveyed sky. We define the Mean Absolute Error Ratio (MAER) as
\begin{equation}
    \mathrm{MAER}^\mathrm{AB}(\boldsymbol{\theta})= \frac{100}{N_{C(\ell)}}\sum_{\ell=\ell_\text{min}}^{\ell_\text{max}}\sum_{i,j}\frac{\bigg | C_{ij,\text{NN}}^\mathrm{AB}-C_{ij,\text{GT}}^\mathrm{AB}\bigg |}{\sigma_{ij}^\mathrm{AB}({\ell})},
    \label{eq:MAER}
\end{equation}
where for clarity, we omitted to write explicitly the dependence on the cosmological parameters $\boldsymbol{\theta}$.

To estimate $\sigma({\ell})$, $f_\mathrm{sky}$ was set to $0.4$ (a value compatible with the $\sim 15,000$ deg$^2$ expected from the Euclid and DESI surveys~\citep{laureijs2011euclid}). 

Equation~\eqref{eq:MAER} does not include the shot-noise contribution, which is a survey-dependent quantity and cannot be rescaled as easily as the (multiplicative) $f_\mathrm{sky}$ term. Neglecting shot noise, however, is a conservative approach: any such term would yield a higher relative performance for our emulators.
\begin{figure}
	\centering
	\includegraphics[width=0.9\linewidth]{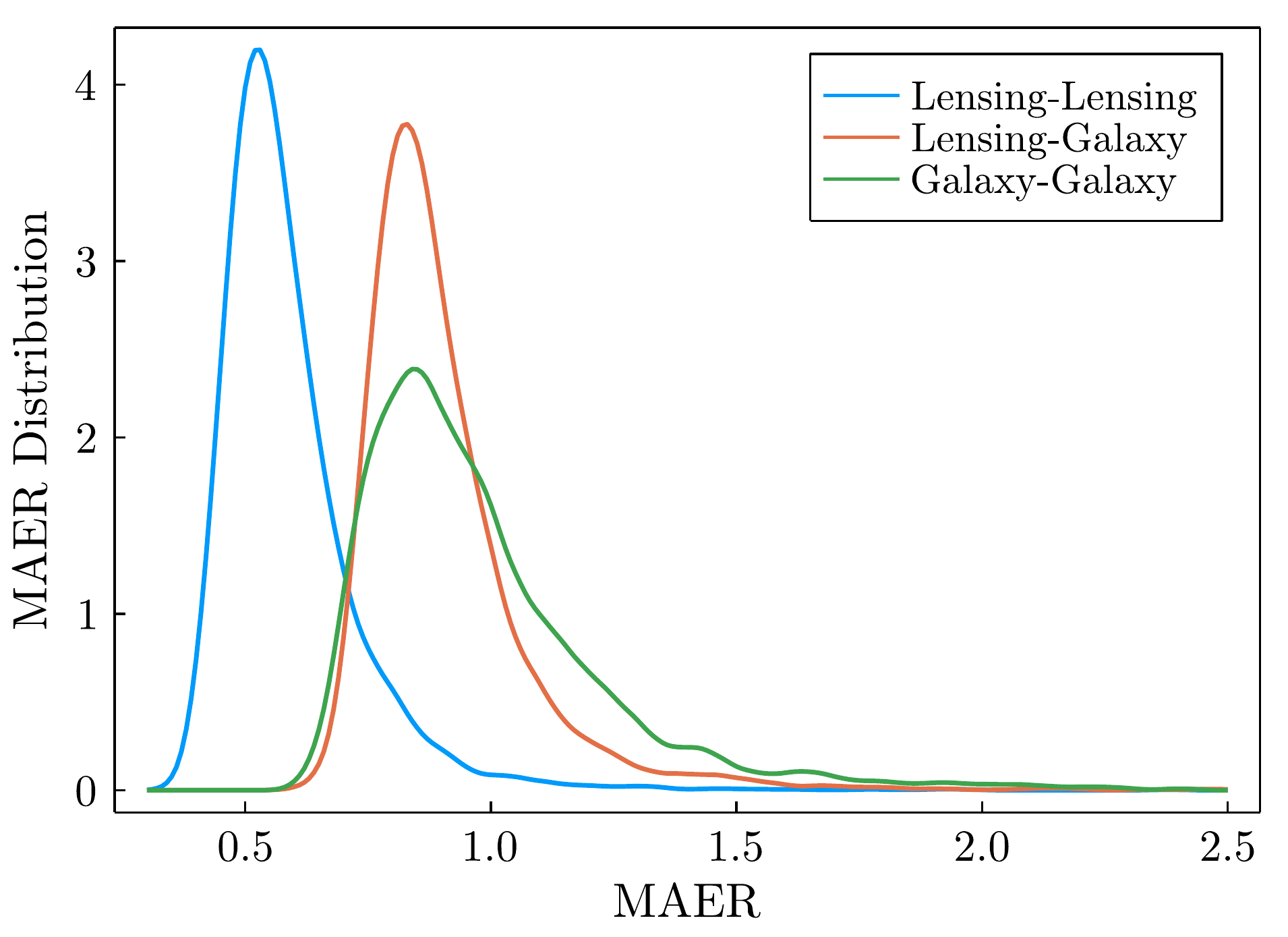}
    \caption{The distribution of the relative importance of systematic errors, with respect to the intrinsic cosmic variance, as defined in Equation~\eqref{eq:MAER}.}
    \label{fig:MAER}
\end{figure}
As can be seen from Figure~\ref{fig:MAER}, the MAER is always smaller than $2\%$. For comparison, the cosmological library developed by the LLST collaboration, \texttt{CCL}~\citep{chisari_core_2019}, set a threshold of $10\%$ for this metric.

\section{Comparison with existing emulators}
\label{sec:comparison}
As we briefly described in the Introduction, there are two categories of LSS emulators in the literature. The first category emulates the matter power spectrum in real space, $P_\mathrm{mm}(k)$, which is then fed to a code that computes the theoretical $C_{ij}^{AB}(\ell)$. The second category \citep{mootoovaloo_parameter_2020, manrique-yus_euclid-era_2019}, instead, emulates directly the latter. Our code belongs to this second class, with our implementation introducing considerable improvement over previous implementation of this kind.

On the other hand, comparison with the first category is more delicate.

For instance, one advantage of $P_\mathrm{mm}(k)$ emulators is that they are more flexible,  since they are agnostic about the galaxy redshift distribution \citep{mancini_itcosmopower_2021}. $C_{ij}^{AB}(\ell)$ emulators, instead, are inevitably tailored for a  specific survey: if the survey changes, the training process needs to be repeated.
However, the most resource-demanding step of our approach is the estimation of the power spectra from Boltzmann codes. Once computed for a large range of redshifts, as done for this work, these could be stored and re-used to evaluate the $C_{ij}^{AB}(\ell)$ with faster codes than current Boltzmann solvers\footnote{For instance, once the power spectrum $P_\mathrm{mm}$ for a given cosmology is provided, \texttt{CosmoCentral.jl} can evaluate the $C_{ij}^{AB}(\ell)$ in about 0.05 s, requiring less than 10 minutes for 10,000 cosmologies.}.
The training procedure can also be accelerated, with \textit{transfer learning} \citep{zhuang_2019}: using a previously trained network as initial ansatz, the network will learn to emulate the new $C_{ij}^{AB}(\ell)$ with less computational resources.

Another advantage of $P_\mathrm{mm}$ emulators is their insensitivity to probe-specific nuisance parameters, which enter the $C_{ij}^{AB}(\ell)$ evaluation through the weight functions $W_i^\mathrm{A}(z)$.
Therefore, the training set in the case of $C_{ij}^{AB}(\ell)$ emulators needs to cover a larger parameter hyperspace.
However, considering that for this work we needed 10,000 cosmologies, the resulting cost is still acceptable and the gain is still positive.  Furthermore, some parameters, as the magnification bias, can be added analytically as multiplicative factors, thus at no cost for the $C_{ij}^{AB}(\ell)$ emulators~\citep{abbott_dark_2018}.

For their part, the $C_{ij}^{AB}(\ell)$ emulators present several advantages. First, they are blazingly fast, since they do not need to compute intermediate quantities like weight functions, background variables and numerical integrals. 
The net result is a gain of about a factor of 1000 in speed, compared to standard Boltzmann integrators.

Emulating directly the $C_{ij}^{AB}(\ell)$, rather than $P_\mathrm{mm}(k)$, becomes even more advantageous when one needs to drop the Limber approximation, when projecting from 3D to 2D statistics. This assumption can in fact become a source of systematic error in high-precision measurements, as expected from next-generation surveys~\citep{fang_beyond_2020, martinelli_ultra-large-scale_2021}. Although efficient algorithms to estimate the $C_{ij}^{AB}(\ell)$ from $P_\mathrm{mm}(k)$ without the Limber approximation  have been recently developed~\citep{schoneberg_beyond_2018, fang_beyond_2020}, the advantage of bypassing completely this computational phase through a direct $C_{ij}^{AB}(\ell)$ emulation, is evident: once the training set of $C_{ij}^{AB}(\ell)$ is built properly, using or not the Limber approximation, the emulator will learn accordingly without any change in the architecture or procedure.
\section{Conclusions}
\label{sec:conclusion}
In this work we have developed a set of fast cosmological emulators, which are capable to predict a specific set of typical 2D statistics from galaxy surveys, the so-called 3x2pt statistics from WL and photometric GC, in less than a millisecond. We have shown that, compared to previous works, they are more precise and flexible, with significant saving of computational and storage resources. 
Precision is also improved, using an appropriate pre-processing step and possibly benefiting of a different architecture: since we are training a single neural network to learn all tomographic combinations, $(i,j)$, at once, the emulator can take advantage of the existing physical correlations between different combinations to deliver the correct result.

Our current results could be further improved by including more astrophysical effects and nuisance parameters as the galaxy bias. 
A full beyond-Limber version of the $C(\ell)$ emulators is also in our plans. As discussed in the previous section, with our approach this is particularly advantageous in terms of computational cost.  We could also further improve in terms of hyperparameters of the neural network: in the current implementation we have used general-purpose dense neural networks, all with the same number of layers and neurons, without further specialization. However, using different architectures for each emulator could increase the precision further. This approach can also be naturally extended to incorporate CMB measurements \citep{mancini_itcosmopower_2021}.

Finally, these techniques would also allow us to improve gradient-based methods in likelihood-based inference approaches: samplers such as the Hamiltonian MonteCarlo~\citep{hoffman_no-u-turn_2011} use the likelihood gradient to efficiently explore the parameter space. However, with standard approaches based on Boltzmann solvers such as \texttt{CAMB} or \texttt{CLASS}~\citep{lewis_efficient_2000, lesgourgues_2011}, evaluating the likelihood gradient  requires computing many finite-difference derivatives, which are both expensive and may suffer poor accuracy and precision. With a neural network the situation is completely different, since by construction the derivative of the loss function with respect to the input is obtained using automatic differentiation. We plan to implement these techniques in a future work.

\section*{Acknowledgements}
MB warmly thanks Luca Paganin, Alessio Spurio Mancini, and Silvio Traversaro for useful discussions. MB acknowledges financial support from the ASI agreement n. I/023/12/0 "Euclid attivitá relativa alla fase B2/C".

\bibliographystyle{mnras}
\bibliography{biblio.bib}

\begin{thebibliography}{}
\makeatletter
\relax
\def\mn@urlcharsother{\let\do\@makeother \do\$\do\&\do\#\do\^\do\_\do\%\do\~}
\def\mn@doi{\begingroup\mn@urlcharsother \@ifnextchar [ {\mn@doi@}
  {\mn@doi@[]}}
\def\mn@doi@[#1]#2{\def\@tempa{#1}\ifx\@tempa\@empty \href
  {http://dx.doi.org/#2} {doi:#2}\else \href {http://dx.doi.org/#2} {#1}\fi
  \endgroup}
\def\mn@eprint#1#2{\mn@eprint@#1:#2::\@nil}
\def\mn@eprint@arXiv#1{\href {http://arxiv.org/abs/#1} {{\tt arXiv:#1}}}
\def\mn@eprint@dblp#1{\href {http://dblp.uni-trier.de/rec/bibtex/#1.xml}
  {dblp:#1}}
\def\mn@eprint@#1:#2:#3:#4\@nil{\def\@tempa {#1}\def\@tempb {#2}\def\@tempc
  {#3}\ifx \@tempc \@empty \let \@tempc \@tempb \let \@tempb \@tempa \fi \ifx
  \@tempb \@empty \def\@tempb {arXiv}\fi \@ifundefined
  {mn@eprint@\@tempb}{\@tempb:\@tempc}{\expandafter \expandafter \csname
  mn@eprint@\@tempb\endcsname \expandafter{\@tempc}}}

\bibitem[\protect\citeauthoryear{Abbott et~al.}{Abbott
  et~al.}{2018a}]{abbott_dark_2018}
Abbott T. M.~C.,  et~al., 2018a, \mn@doi [Phys. Rev. D]
  {10.1103/PhysRevD.98.043526}, 98, 043526

\bibitem[\protect\citeauthoryear{Abbott et~al.}{Abbott
  et~al.}{2018b}]{abbott_dark_2018-1}
Abbott T. M.~C.,  et~al., 2018b, \mn@doi [Astrophys. J. Suppl.]
  {10.3847/1538-4365/aae9f0}, 239, 18

\bibitem[\protect\citeauthoryear{Aghanim et~al.}{Aghanim
  et~al.}{2020}]{aghanim_planck_2020}
Aghanim N.,  et~al., 2020, \mn@doi [Astron. Astrophys.]
  {10.1051/0004-6361/201833910}, 641, A6

\bibitem[\protect\citeauthoryear{Albers, Fidler, Lesgourgues, Sch\"oneberg  \&
  Torrado}{Albers et~al.}{2019}]{albers_cosmicnet_2019}
Albers J.,  Fidler C.,  Lesgourgues J.,  Sch\"oneberg N.,   Torrado J.,  2019,
  \mn@doi [JCAP] {10.1088/1475-7516/2019/09/028}, 09, 028

\bibitem[\protect\citeauthoryear{Angulo, Zennaro, Contreras, Aric\`o,
  Pellejero-Iba\~nez  \& St\"ucker}{Angulo et~al.}{2021}]{angulo_bacco_2021}
Angulo R.~E.,  Zennaro M.,  Contreras S.,  Aric\`o G.,  Pellejero-Iba\~nez M.,
   St\"ucker J.,  2021, \mn@doi [Mon. Not. Roy. Astron. Soc.]
  {10.1093/mnras/stab2018}, 507, 5869

\bibitem[\protect\citeauthoryear{{Aric{\`o}}, {Angulo}  \&
  {Zennaro}}{{Aric{\`o}} et~al.}{2021}]{arico_accelerating_2021}
{Aric{\`o}} G.,  {Angulo} R.~E.,   {Zennaro} M.,  2021, arXiv e-prints, \href
  {https://ui.adsabs.harvard.edu/abs/2021arXiv210414568A} {p. arXiv:2104.14568}

\bibitem[\protect\citeauthoryear{Bernal, Verde  \& Riess}{Bernal
  et~al.}{2016}]{bernal_trouble_2016}
Bernal J.~L.,  Verde L.,   Riess A.~G.,  2016, \mn@doi [JCAP]
  {10.1088/1475-7516/2016/10/019}, 10, 019

\bibitem[\protect\citeauthoryear{Blanchard et~al.}{Blanchard
  et~al.}{2020}]{blanchard_euclid_2020}
Blanchard A.,  et~al., 2020, \mn@doi [Astron. Astrophys.]
  {10.1051/0004-6361/202038071}, 642, A191

\bibitem[\protect\citeauthoryear{Brinckmann \& Lesgourgues}{Brinckmann \&
  Lesgourgues}{2019}]{brinckmann_montepython_2018}
Brinckmann T.,  Lesgourgues J.,  2019, \mn@doi [Phys. Dark Univ.]
  {10.1016/j.dark.2018.100260}, 24, 100260

\bibitem[\protect\citeauthoryear{Carrasco, Hertzberg  \& Senatore}{Carrasco
  et~al.}{2012}]{carrasco_effective_2012}
Carrasco J. J.~M.,  Hertzberg M.~P.,   Senatore L.,  2012, \mn@doi [JHEP]
  {10.1007/JHEP09(2012)082}, 09, 082

\bibitem[\protect\citeauthoryear{Cataneo, Lombriser, Heymans, Mead, Barreira,
  Bose  \& Li}{Cataneo et~al.}{2019}]{cataneo_road_2019}
Cataneo M.,  Lombriser L.,  Heymans C.,  Mead A.,  Barreira A.,  Bose S.,   Li
  B.,  2019, \mn@doi [Mon. Not. Roy. Astron. Soc.] {10.1093/mnras/stz1836},
  488, 2121

\bibitem[\protect\citeauthoryear{Chevallier \& Polarski}{Chevallier \&
  Polarski}{2001}]{chevallier_accelerating_2001}
Chevallier M.,  Polarski D.,  2001, \mn@doi [Int. J. Mod. Phys. D]
  {10.1142/S0218271801000822}, 10, 213

\bibitem[\protect\citeauthoryear{Chisari et~al.}{Chisari
  et~al.}{2019}]{chisari_core_2019}
Chisari N.~E.,  et~al., 2019, \mn@doi [Astrophys. J. Suppl.]
  {10.3847/1538-4365/ab1658}, 242, 2

\bibitem[\protect\citeauthoryear{Cybenko}{Cybenko}{1989}]{cybenko_approximation_1989}
Cybenko G.,  1989, \mn@doi [Math. Control Signal Systems] {10.1007/BF02551274},
  2, 303

\bibitem[\protect\citeauthoryear{Desjacques, Jeong  \& Schmidt}{Desjacques
  et~al.}{2018}]{desjacques_large-scale_2018}
Desjacques V.,  Jeong D.,   Schmidt F.,  2018, \mn@doi [Phys. Rept.]
  {10.1016/j.physrep.2017.12.002}, 733, 1

\bibitem[\protect\citeauthoryear{{Dore} et~al.,}{{Dore}
  et~al.}{2019}]{dore_wfirst_2019}
{Dore} O.,  et~al., 2019, \baas, \href
  {https://ui.adsabs.harvard.edu/abs/2019BAAS...51c.341D} {51, 341}

\bibitem[\protect\citeauthoryear{Fang, Krause, Eifler  \& MacCrann}{Fang
  et~al.}{2020}]{fang_beyond_2020}
Fang X.,  Krause E.,  Eifler T.,   MacCrann N.,  2020, \mn@doi [JCAP]
  {10.1088/1475-7516/2020/05/010}, 05, 010

\bibitem[\protect\citeauthoryear{Feldman, Kaiser  \& Peacock}{Feldman
  et~al.}{1994}]{feldman_power_1994}
Feldman H.~A.,  Kaiser N.,   Peacock J.~A.,  1994, \mn@doi [Astrophys. J.]
  {10.1086/174036}, 426, 23

\bibitem[\protect\citeauthoryear{Fendt \& Wandelt}{Fendt \&
  Wandelt}{2006}]{fendt_pico_2007}
Fendt W.~A.,  Wandelt B.~D.,  2006, \mn@doi [Astrophys. J.] {10.1086/508342},
  654, 2

\bibitem[\protect\citeauthoryear{Hoffman \& Gelman}{Hoffman \&
  Gelman}{2014}]{hoffman_no-u-turn_2011}
Hoffman M.~D.,  Gelman A.,  2014, J. Mach. Learn. Res., 15, 1593

\bibitem[\protect\citeauthoryear{Hornik, Stinchcombe  \& White}{Hornik
  et~al.}{1989}]{hornik_multilayer_1989}
Hornik K.,  Stinchcombe M.,   White H.,  1989, \mn@doi [Neural Networks]
  {10.1016/0893-6080(89)90020-8}, 2, 359

\bibitem[\protect\citeauthoryear{Hu}{Hu}{1999}]{hu_power_1999}
Hu W.,  1999, \mn@doi [Astrophys. J. Lett.] {10.1086/312210}, 522, L21

\bibitem[\protect\citeauthoryear{{Innes} et~al.,}{{Innes}
  et~al.}{2018}]{innes_fashionable_2018}
{Innes} M.,  et~al., 2018, arXiv e-prints, \href
  {https://ui.adsabs.harvard.edu/abs/2018arXiv181101457I} {p. arXiv:1811.01457}

\bibitem[\protect\citeauthoryear{Ioffe \& Szegedy}{Ioffe \&
  Szegedy}{2015}]{ioffe_batch_2015}
Ioffe S.,  Szegedy C.,  2015, arXiv:1502.03167 [cs]

\bibitem[\protect\citeauthoryear{Jimenez, Verde, Peiris  \& Kosowsky}{Jimenez
  et~al.}{2004}]{jimenez_fast_2004}
Jimenez R.,  Verde L.,  Peiris H.,   Kosowsky A.,  2004, \mn@doi [Phys. Rev. D]
  {10.1103/PhysRevD.70.023005}, 70, 023005

\bibitem[\protect\citeauthoryear{{Kingma} \& {Ba}}{{Kingma} \&
  {Ba}}{2014}]{kingma_adam}
{Kingma} D.~P.,  {Ba} J.,  2014, arXiv e-prints, \href
  {https://ui.adsabs.harvard.edu/abs/2014arXiv1412.6980K} {p. arXiv:1412.6980}

\bibitem[\protect\citeauthoryear{Kitching, Amara, Abdalla, Joachimi  \&
  Refregier}{Kitching et~al.}{2009}]{kitching_cosmological_2009}
Kitching T.~D.,  Amara A.,  Abdalla F.~B.,  Joachimi B.,   Refregier A.,  2009,
  \mn@doi [Mon. Not. Roy. Astron. Soc.] {10.1111/j.1365-2966.2009.15408.x},
  399, 2107

\bibitem[\protect\citeauthoryear{Knabenhans et~al.}{Knabenhans
  et~al.}{2021}]{euclidcollaboration_euclid_2021}
Knabenhans M.,  et~al., 2021, \mn@doi [Mon. Not. Roy. Astron. Soc.]
  {10.1093/mnras/stab1366}, 505, 2840

\bibitem[\protect\citeauthoryear{{Laureijs} et~al.,}{{Laureijs}
  et~al.}{2011a}]{laureijs_euclid_2011}
{Laureijs} R.,  et~al., 2011a, arXiv e-prints, \href
  {https://ui.adsabs.harvard.edu/abs/2011arXiv1110.3193L} {p. arXiv:1110.3193}

\bibitem[\protect\citeauthoryear{{Laureijs} et~al.,}{{Laureijs}
  et~al.}{2011b}]{laureijs2011euclid}
{Laureijs} R.,  et~al., 2011b, arXiv e-prints, \href
  {https://ui.adsabs.harvard.edu/abs/2011arXiv1110.3193L} {p. arXiv:1110.3193}

\bibitem[\protect\citeauthoryear{{Lesgourgues}}{{Lesgourgues}}{2011}]{lesgourgues_2011}
{Lesgourgues} J.,  2011, arXiv e-prints, \href
  {https://ui.adsabs.harvard.edu/abs/2011arXiv1104.2932L} {p. arXiv:1104.2932}

\bibitem[\protect\citeauthoryear{Lewis, Challinor  \& Lasenby}{Lewis
  et~al.}{2000}]{lewis_efficient_2000}
Lewis A.,  Challinor A.,   Lasenby A.,  2000, \mn@doi [Astrophys. J.]
  {10.1086/309179}, 538, 473

\bibitem[\protect\citeauthoryear{Limber}{Limber}{1954}]{limber_analysis_1953}
Limber D.~N.,  1954, \mn@doi [Astrophys. J.] {10.1086/145870}, 119, 655

\bibitem[\protect\citeauthoryear{Linder}{Linder}{2003}]{linder_exploring_2003}
Linder E.~V.,  2003, \mn@doi [Phys. Rev. Lett.]
  {10.1103/PhysRevLett.90.091301}, 90, 091301

\bibitem[\protect\citeauthoryear{LoVerde \& Afshordi}{LoVerde \&
  Afshordi}{2008}]{loverde_extended_2008}
LoVerde M.,  Afshordi N.,  2008, \mn@doi [Phys. Rev. D]
  {10.1103/PhysRevD.78.123506}, 78, 123506

\bibitem[\protect\citeauthoryear{Manrique-Yus \& Sellentin}{Manrique-Yus \&
  Sellentin}{2020}]{manrique-yus_euclid-era_2019}
Manrique-Yus A.,  Sellentin E.,  2020, \mn@doi [Mon. Not. Roy. Astron. Soc.]
  {10.1093/mnras/stz3059}, 491, 2655

\bibitem[\protect\citeauthoryear{Martinelli, Dalal, Majidi, Akrami, Camera  \&
  Sellentin}{Martinelli et~al.}{2022}]{martinelli_ultra-large-scale_2021}
Martinelli M.,  Dalal R.,  Majidi F.,  Akrami Y.,  Camera S.,   Sellentin E.,
  2022, \mn@doi [Mon. Not. Roy. Astron. Soc.] {10.1093/mnras/stab3578}, 510,
  1964

\bibitem[\protect\citeauthoryear{McKay, Beckman  \& Conover}{McKay
  et~al.}{1979}]{latin_hypercube}
McKay M.~D.,  Beckman R.~J.,   Conover W.~J.,  1979, Technometrics, 21, 239

\bibitem[\protect\citeauthoryear{Mootoovaloo, Heavens, Jaffe  \&
  Leclercq}{Mootoovaloo et~al.}{2020}]{mootoovaloo_parameter_2020}
Mootoovaloo A.,  Heavens A.~F.,  Jaffe A.~H.,   Leclercq F.,  2020, \mn@doi
  [Mon. Not. Roy. Astron. Soc.] {10.1093/mnras/staa2102}, 497, 2213

\bibitem[\protect\citeauthoryear{Mootoovaloo, Jaffe, Heavens  \&
  Leclercq}{Mootoovaloo et~al.}{2022}]{mootoovaloo_kernel-based_2021}
Mootoovaloo A.,  Jaffe A.~H.,  Heavens A.~F.,   Leclercq F.,  2022, \mn@doi
  [Astron. Comput.] {10.1016/j.ascom.2021.100508}, 38, 100508

\bibitem[\protect\citeauthoryear{{Nwankpa}, {Ijomah}, {Gachagan}  \&
  {Marshall}}{{Nwankpa} et~al.}{2018}]{nwankpa_relu}
{Nwankpa} C.,  {Ijomah} W.,  {Gachagan} A.,   {Marshall} S.,  2018, arXiv
  e-prints, \href {https://ui.adsabs.harvard.edu/abs/2018arXiv181103378N} {p.
  arXiv:1811.03378}

\bibitem[\protect\citeauthoryear{Perlmutter et~al.}{Perlmutter
  et~al.}{1999}]{perlmutter_measurements_1999}
Perlmutter S.,  et~al., 1999, \mn@doi [Astrophys. J.] {10.1086/307221}, 517,
  565

\bibitem[\protect\citeauthoryear{Riess et~al.}{Riess
  et~al.}{1998}]{riess_observational_1998}
Riess A.~G.,  et~al., 1998, \mn@doi [Astron. J.] {10.1086/300499}, 116, 1009

\bibitem[\protect\citeauthoryear{Sch\"oneberg, Simonovi\'c, Lesgourgues  \&
  Zaldarriaga}{Sch\"oneberg et~al.}{2018}]{schoneberg_beyond_2018}
Sch\"oneberg N.,  Simonovi\'c M.,  Lesgourgues J.,   Zaldarriaga M.,  2018,
  \mn@doi [JCAP] {10.1088/1475-7516/2018/10/047}, 10, 047

\bibitem[\protect\citeauthoryear{Spurio~Mancini, Piras, Alsing, Joachimi  \&
  Hobson}{Spurio~Mancini et~al.}{2022}]{mancini_itcosmopower_2021}
Spurio~Mancini A.,  Piras D.,  Alsing J.,  Joachimi B.,   Hobson M.~P.,  2022,
  \mn@doi [Mon. Not. Roy. Astron. Soc.] {10.1093/mnras/stac064}, 511, 1771

\bibitem[\protect\citeauthoryear{Takahashi, Sato, Nishimichi, Taruya  \&
  Oguri}{Takahashi et~al.}{2012}]{takahashi_revising_2012}
Takahashi R.,  Sato M.,  Nishimichi T.,  Taruya A.,   Oguri M.,  2012, \mn@doi
  [Astrophys. J.] {10.1088/0004-637X/761/2/152}, 761, 152

\bibitem[\protect\citeauthoryear{Torrado \& Lewis}{Torrado \&
  Lewis}{2021}]{torrado_cobaya_2021}
Torrado J.,  Lewis A.,  2021, \mn@doi [JCAP] {10.1088/1475-7516/2021/05/057},
  05, 057

\bibitem[\protect\citeauthoryear{Tutusaus et~al.}{Tutusaus
  et~al.}{2020}]{tutusaus_euclid_2020}
Tutusaus I.,  et~al., 2020, \mn@doi [Astron. Astrophys.]
  {10.1051/0004-6361/202038313}, 643, A70

\bibitem[\protect\citeauthoryear{Weltman et~al.}{Weltman
  et~al.}{2020}]{weltman_fundamental_2020}
Weltman A.,  et~al., 2020, \mn@doi [Publ. Astron. Soc. Austral.]
  {10.1017/pasa.2019.42}, 37, e002

\bibitem[\protect\citeauthoryear{Yamamoto, Nakamichi, Kamino, Bassett  \&
  Nishioka}{Yamamoto et~al.}{2006}]{yamamoto_measurement_2006}
Yamamoto K.,  Nakamichi M.,  Kamino A.,  Bassett B.~A.,   Nishioka H.,  2006,
  \mn@doi [Publ. Astron. Soc. Jap.] {10.1093/pasj/58.1.93}, 58, 93

\bibitem[\protect\citeauthoryear{{Zhuang}, {Qi}, {Duan}, {Xi}, {Zhu}, {Zhu},
  {Xiong}  \& {He}}{{Zhuang} et~al.}{2019}]{zhuang_2019}
{Zhuang} F.,  {Qi} Z.,  {Duan} K.,  {Xi} D.,  {Zhu} Y.,  {Zhu} H.,  {Xiong} H.,
    {He} Q.,  2019, arXiv e-prints, \href
  {https://ui.adsabs.harvard.edu/abs/2019arXiv191102685Z} {p. arXiv:1911.02685}

\makeatother
\end{thebibliography}

\end{document}